\documentclass[a4paper]{article}

\usepackage{INTERSPEECH2022}
\usepackage[table, svgnames, dvipsnames]{xcolor}
\usepackage{hyperref}
\usepackage[normalem]{ulem}
\definecolor{mygreen}{HTML}{008000}

\title{MAE-AST: Masked Autoencoding Audio Spectrogram Transformer}
\name{Alan Baade, Puyuan Peng, David Harwath}
\address{Department of Computer Science, The University of Texas at Austin}
\email{abaade,pyp,harwath@utexas.edu}

\begin{document}

\maketitle
\begin{abstract}
In this paper, we propose a simple yet powerful improvement over the recent Self-Supervised Audio Spectrogram Transformer (SSAST) model for speech and audio classification. Specifically, we leverage the insight that the SSAST uses a very high masking ratio (75\%) during pretraining, meaning that the vast majority of self-attention compute is performed on mask tokens. We address this by integrating the encoder-decoder architecture from Masked Autoencoders are Scalable Vision Learners (MAE) into the SSAST, where a deep encoder operates on only unmasked input, and a shallow decoder operates on encoder outputs and mask tokens. We find that MAE-like pretraining can provide a 3× speedup and 2× memory usage reduction over the vanilla SSAST using current audio pretraining strategies with ordinary model and input sizes. When fine-tuning on downstream tasks, which only uses the encoder, we find that our approach outperforms the SSAST on a variety of downstream tasks. We further conduct comprehensive evaluations into different strategies of pretraining and explore differences in MAE-style pretraining between the visual and audio domains.



\end{abstract}

\section{Introduction and Related Work}
Training high-quality neural networks requires lots of data. However, simple-to-train-with labeled data is expensive and limited, while unlabeled data from the internet is far more abundant. To address this, self-supervised learning techniques have seen an explosion in research interest, especially after producing incredible results in the natural language domain~\cite{devlin2018bert,radford2019language,brown2020language}.

Recently, the success of self-supervised pretraining has migrated into the audio~\cite{baevski2020wav2vec,hsu2021hubert} and vision~\cite{DBLP:journals/corr/abs-2010-11929,liu2021Swin} communities. Of particular interest to this paper is the Self-Supervised Audio Spectrogram Transformer (SSAST)~\cite{gong2021ssast}, a self-supervised extension to the Audio Spectrogram Transformer (AST)~\cite{gong21b_interspeech}. The SSAST introduced the first patch-based and fully self-attention~\cite{DBLP:journals/corr/VaswaniSPUJGKP17} based pretraining strategies for audio event classification, and achieved remarkable results for both audio event classification and several speech classification tasks. However, one common criticism of Transformer models is their massive computational overhead, in large part due to their $O(N^2)$ memory and time complexity when computing the self-attention operation. Many so-called ``efficient'' Transformers have been proposed~\cite{Tay2020EfficientTA} that try to mitigate this cost. Of particular relevence to us is the Masked Autoencoder~\cite{MaskedAutoencoders2021}, which completely discards masked input tokens during the encoding step, resulting in significant increases in computational efficiency.

In this paper, we improve upon the SSAST architecture by incorporating ideas from the Masked Autoencoder (MAE) introduced by Kaiming et al.~\cite{MaskedAutoencoders2021}, which asymmetrically applies BERT-like~\cite{devlin2018bert} pretraining to the visual domain with an encoder-decoder architecture. Specifically, we observe that we can reduce the input length to a majority of transformer layers by taking advantage of the high masking ratio used for audio pretraining (75\%). We additionally notice that, similar to the visual domain, reconstruction in the audio spectrogram domain introduces generally unwanted, low-level pixel information to the final layers of pretrained models. From these observations, we propose the Masked Autoencoding Audio Spectrogram Transformer (MAE-AST), which runs only unmasked tokens through a large encoder and then concatenates mask tokens with encoder output embeddings before feeding them to a shallow decoder. Fine-tuning on downstream tasks is then performed using only the encoder, removing the decoder's reconstruction layers.

We find two key results: 1) The MAE-AST pretrains using significantly less time ($3\times$) and memory ($2\times$) than the SSAST for a similar model architecture. 2) The MAE-AST outperforms the SSAST under a shared encoder depth on several downstream tasks, all other factors held constant. Additionally, we find that the MAE-AST performs well with only a generative objective, while the SSAST sees a significant performance drop. We further run several experiments on methods to pretrain the MAE-AST and compare MAE-like pretraining on audio to both SSAST-like pretraining on audio and MAE-like pretraining on vision.

\section{Masked Autoencoding Audio Spectrogram Transformer}

\begin{figure}[t]
  \centering
  \includegraphics[width=\linewidth]{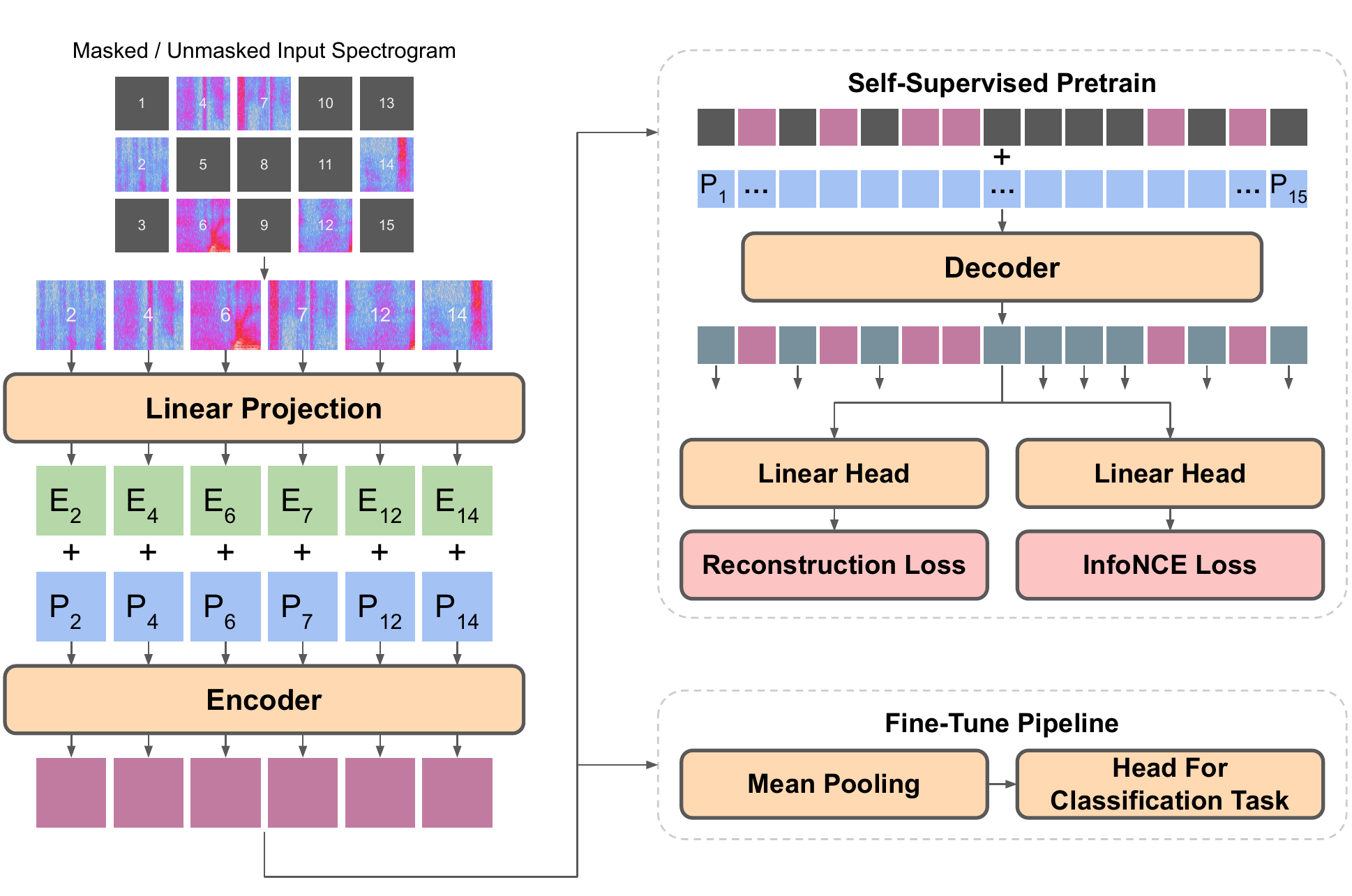}
  \caption{The architecture of the  MAE-AST for patch-based inputs. First, the input spectrogram of size $128\times 100t$ is split into $16\times 16$ pixel patches. For pretraining, a mask is then applied to some tokens. Unmasked patches are unrolled first by channel dimension and then by time dimension and are embedded into a vector via linear projection. These unmasked patch embeddings are added to 1D sinusoidal positional embeddings and fed into the encoder. During pretraining, mask tokens are combined with encoder outputs and positional embeddings are added to all tokens before being sent through the decoder. The joint discriminative and generative loss is calculated on the decoder outputs for mask tokens. For fine-tuning, the encoder outputs are mean pooled to create a representation vector and the decoder isn’t touched.}
  \label{fig:model_diagram}
\end{figure}

Figure~\ref{fig:model_diagram} shows the  MAE-AST architecture. In this section, we first outline the forward pass, followed by training practices, masking strategies, and the loss function.

\subsection{Model Architecture}

\textbf{Input Processing.} Identical to the SSAST, we take as input a 16khz audio waveform and convert it into 128-dimensional log Mel filterbank features with a frame length of 25 ms and frame shift of 10 ms. Like AST and SSAST, we normalize the input to mean 0 and standard deviation ½ . This spectrogram is then split when patch-based into a series of 16 filter $\times$ 16 frame tokens and when frame-based into 128 filter $\times$ 2 frame tokens. We then apply a masking method, which splits the input patches or frames into masked and unmasked subsets. We experiment with several masking strategies, discussed later.

\textbf{Positional Embeddings.} Consistent with~\cite{MaskedAutoencoders2021}, we use fixed sinusoidal positional embeddings for both patch-based and frame-based tokenization. As the input spectrogram data is only variable-length in time, we use one-dimensional positional embeddings for both tokenizations, with patch input being unfolded into an one dimensional stream ordered first by channel and second by time to support variable input lengths, as shown in Figure~\ref{fig:model_diagram}.

\textbf{Encoder.} The encoder follows a standard ViT~\cite{DBLP:journals/corr/abs-2010-11929} architecture. Unlike BERT-like models, the MAE encoder only operates on unmasked tokens. We flatten the unmasked input patches or frames via a linear projection into a 768-dimensional latent vector. We then add positional embeddings to these latents and input the summed vectors to the encoder.

\textbf{Decoder.} The decoder is only used during pretraining and is composed of standard transformer encoder blocks. Both the encoder output and mask tokens are sent to the decoder, with mask tokens at the decoder's input being represented by a shared, learned embedding. We add sinusoidal positional embeddings to all tokens before inputting to the decoder. The output embeddings of masked tokens from the decoder are then sent into two single-layer linear projections, one for each loss function, that map to a vector the size of the patch input. Specifics of calculating loss are discussed later.

By default, we use an encoder with 6 layers, and a decoder of 2 layers. Both the encoder and decoder use 12 heads and a width of 768. Following the MAE~\cite{MaskedAutoencoders2021}, by default we throw away the decoder during fine-tuning, training solely with the encoder. We mean pool the encoder output last hidden states to create a single vector representation for classification tasks.

To create a fair comparison between MAE-like and traditional BERT-like architectures for audio pretraining, we largely adopt the input pipelining and loss calculation techniques from the SSAST. However, we make two minor changes:

1) During training, the original AST paper~\cite{gong21b_interspeech} applies overlap to its input patches for a modest performance improvement. The SSAST paper does not apply overlap during pretraining to prevent the model from using it to cheat the reconstruction task. However, SSAST adds overlap back during fine-tuning to match the original AST paper via interpolating learned positional embeddings. To match the MAE paper, we forgo overlap during both pretraining and fine-tuning.

2) The SSAST uses interpolation or truncation of positional embeddings to support variable-length inputs during fine-tuning. To match the MAE paper, we make no modifications to our positional embeddings for fine-tuning.

We perform experiments under the fairseq library~\cite{ott2019fairseq} using the Adam~\cite{adam} optimizer with weight decay 0.01, initial learning rate 0.0001, and a polynomial decay learning rate scheduler. With fairseq, we use a maximum number of tokens per GPU that is roughly equivalent to a batch size of 32. Training runs for up to 600,000 iterations, which provides the same number of audio clips during pretraining as SSAST. We train using one NVIDIA Quadro RTX-8000, and pretraining takes about two days.

\subsection{Mask Sampling}
Mask sampling for pretraining generally involves two considerations, the percent of tokens to mask and the amount of chunking between masked tokens. Masking strategies with too little chunking or too few masked tokens will create too easy of a task, allowing the model to simply interpolate missing pixels using their neighbors. The proportion to mask is especially important for MAE-based models, as a higher masking ratio decreases the length of the input to the decoder, which can provide significant speedups due to the quadratic time and memory complexity of Transformers.

We experiment with eight overall masking and input strategies via combinations of the following factors: 1) Patch-Based versus Frame-Based inputs, 2) Fully random versus chunked masking, 3) Different masking ratios. We implement these strategies and relate them to previous work as follows:

1) Patch-Based, Fully Random masking is the strategy implemented by the original MAE paper for image pretraining. Like the MAE paper, we mask by shuffling the input patches and keeping the first $1-p$ proportion of tokens. 2) Patch-Based, Chunked masking is adapted using the strategy from SSAST, which selects a level of chunking $C \in \{3,4,5\}$ and randomly masks $C \times C$ chunks of tokens until a proportion $\ge p$ is masked, then truncating down to $p$. This unbiased sampling is slow, so we apply the same mask across a batch to reduce compute. 3) Frame-Based, Fully Random masking is the strategy implemented by SSAST for frame-based experiments, and mask selection is performed in the same way as patch-based, fully random masking. 4) Frame-Based, Chunked masking is implemented using the algorithm from Wav2Vec2.0~\cite{baevski2020wav2vec}, where spans of length $M=10$ are masked with each index having a $P$ probability of being chosen to start a span. This method often creates overlap between chunks, making the output length unpredictable. To make fair comparisons between different masking strategies, we select $P$ so that the average proportion of tokens masked is $p$ after accounting for overlap.

\subsection{Joint Discriminative and Generative Pretraining}
One of the key contributions of the SSAST paper was combining both discriminative (classification) and generative (reconstruction) losses for audio pretraining. Our model continues to adapt this strategy, however with a few small differences.

In the SSAST, the encoder output is fed into two separate two-layer feed forward neural networks for self-supervised pretraining, one each for reconstruction and classification loss. Because the decoder head of MAE-based models already provides compute past the last layer of embeddings used, we use a single linear layer head for our final output projection for each loss.

We implement loss functions identically to SSAST. Reconstruction loss is calculated as a mean-squared error between the unnormalized output of the linear reconstruction head and the normalized input. Classification loss intends to create a vector $c_i$ similar to the masked input patch or frame $x_i$ but dissimilar to other masked input. Like the SSAST, we obtain negative samples from other masked inputs within the same audio segment, which both helps lessen useless features being encoded (such as microphone type) and reduces the impact of batch size on pretraining. We use the InfoNCE~\cite{oord2018representation} loss for classification. The classification and reconstruction losses are summed and balanced by a factor $\lambda$ which is multiplied to the reconstruction loss. We use the same $\lambda = 10$ as SSAST.




\section{Experiments}
\subsection{Datasets}

\textbf{Pretraining Datasets.} To fairly show that the MAE encoder-decoder architecture is the cause of performance gains, we use the same pretraining datasets as SSAST. The SSAST pretraining data consists of uniformly shuffled data from AudioSet-2M~\cite{gemmeke2017audio} and Librispeech~\cite{panayotov2015librispeech} with labels discarded. AudioSet-2M contains two million ten second audio clips from YouTube, and provides data for both audio events and speech. AudioSet is subdivided into unbalanced, balanced, and eval partitions, with the balanced and eval sections having roughly the same number of examples per classification label. Librispeech provides 960 hours of audiobook data and is included because AudioSet clips do not necessarily contain speech. We pretrain using both the unbalanced and balanced AudioSet partitions in addition to the librispeech train data.

\textbf{Fine-Tuning Datasets.} We fine-tune on the same six commonly-used datasets as SSAST, which provide a variety of downstream audio and speech challenges. We evaluate audio event classification performance by fine-tuning on the AudioSet (AS) and ESC-50~\cite{piczak2015esc} (ESC) datasets. AudioSet provides a general audio classification task, while ESC-50 tests classification of environmental audio. Like the SSAST paper, we fine-tune on only the balanced partition of AudioSet and test on the eval partition. We evaluate speech performance by fine-tuning on Speech Commands 1 and 2~\cite{warden2018speech} for the task of keyword spotting (KS1, KS2), VoxCeleb~\cite{nagrani2020voxceleb} for speaker identification (SID), and IEMOCAP~\cite{busso2008iemocap} for emotion recognition (ER). For AS, ESC, and KS2, we use the fine-tuning pipeline from SSAST. For KS1, SID, and ER we use the SUPERB~\cite{yang2021superb} framework, the same fine-tune setting as SSAST. We use default settings for all models, which matched the results of a learning rate gridsearch.

\textbf{AudioSet Preprocessing} Because the AudioSet data is downloaded from YouTube directly, videos get deleted and the available dataset decreases in size over time: Our downloaded copy of AudioSet contains 86.4\% of the original dataset, while SSAST had 94\%. To make our results roughly comparable to other papers, we fill in the missing data per label for the balanced and eval sets by sampling from the unbalanced set using the same sampling method as the AudioSet paper. Assuming no correlation between deleted videos and label type, this acts as a random resample of the balanced and eval sets from all downloaded data given that the available balanced and eval sets are chosen, which is the best we can do with limited data. After resampling, our unbalanced set contains 1,762,073 examples, balanced 22,680, and eval 20,920.

We load pretrained weights of SSAST distributed by \cite{gong2021ssast} and fine-tune and test it on our version of AudioSet, meaning all experiments are evaluated on the same data. Importantly, we do not replicate pretraining. However, this should only serve as a disadvantage to our models as recently downloaded data is a subset of data from earlier downloads, barring a small amount of reuploaded videos.

\subsection{Results}

\textbf{Overall Results.} Table \ref{tab:overall_results} shows the results of the MAE-AST with different encoder depths against the SSAST with equivalent input processing and masking techniques. We find that the MAE-AST matches or outperforms the SSAST on all downstream tasks but SID. We fail to make the 12-Layer Patch-Based MAE-AST converge on SID, and further discuss SID later.

\begin{figure}[t]
  \centering
  \includegraphics[width=\linewidth]{ 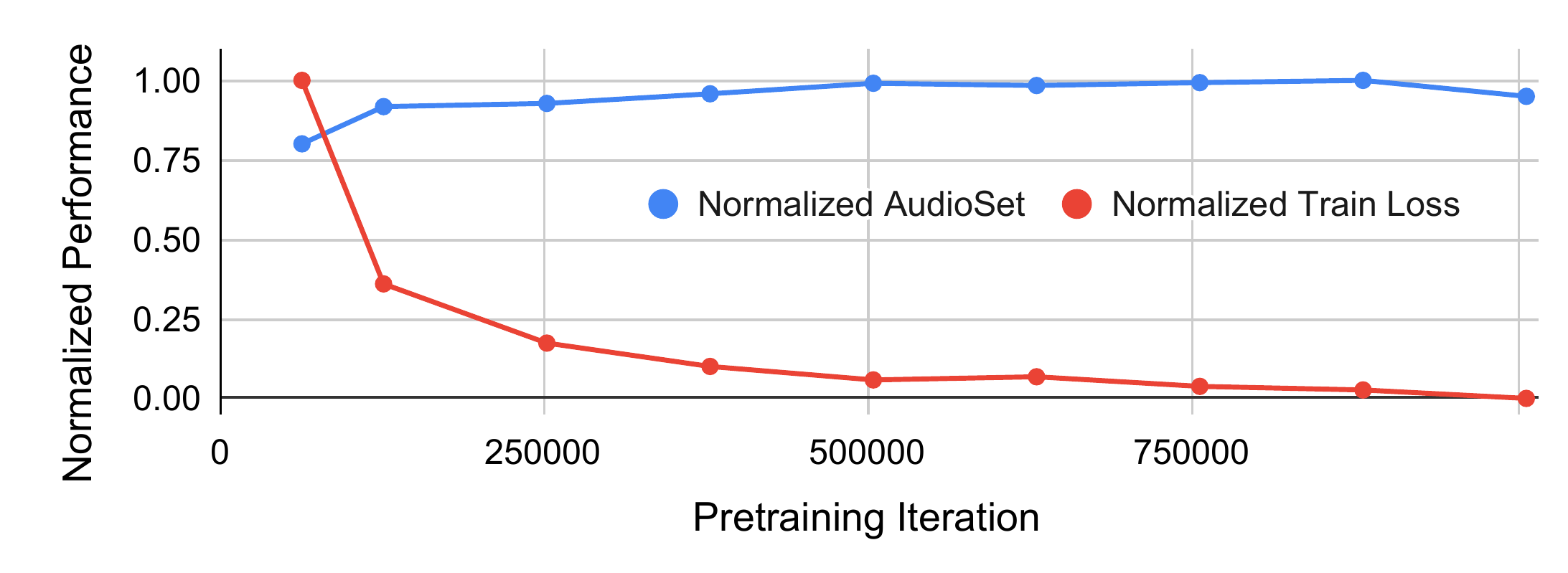}
  \caption{Convergence of MAE-AST. We train the default model for an extended amount and save a checkpoint at training epochs 1,2,4,6,8,10,12,14, and 16. We evaluate these saves on the downstream AudioSet task, and normalize the results in [0,1]. We find that  MAE-AST pretraining converges in 8 epochs (500,000 iterations), the exact same as SSAST.  MAE-AST is also non-brittle to increased pretraining time, maintaining performance after pretraining for twice as long as initial convergence.}
  \label{fig:convergence_chart}
\end{figure}

\begin{table*}[th]
  \caption{Overall Results. We do not replicate SSAST Base Frame on AudioSet for fairness because the exact fine-tune parameters aren't published. Note that the numbers we report for the SSAST model differ from those reported by~\cite{gong2021ssast}. This is due to the fact that we train and test all models on our modified version of the AudioSet dataset.}
  \label{tab:overall_results}
  \centering
  \begin{tabular}{ l c l c c c c c c }
    \toprule
    \multicolumn{1}{l}{\textbf{Model}} & 
    \multicolumn{1}{c}{\textbf{Enc. Layers}} & 
    \multicolumn{1}{l}{\textbf{Masking}} & 
    \multicolumn{1}{c}{\textbf{AS}} & 
    \multicolumn{1}{c}{\textbf{ESC}} & 
    \multicolumn{1}{c}{\textbf{KS2}} & 
    \multicolumn{1}{c}{\textbf{KS1}} & 
    \multicolumn{1}{c}{\textbf{SID}} & 
    \multicolumn{1}{c}{\textbf{ER}} \\
    \midrule
SSAST Base Patch & 12       & Chunked & 28.6    & 88.8    & 98.0     & 96.0     & 64.3     & 59.6    \\
SSAST Base Frame & 12       & Random  & \multicolumn{0}{c}{-}  & 85.9    & \textbf{98.1}     & 96.7     & \textbf{80.8}     & 60.5    \\
 MAE-AST Patch     & 6        & Chunked & 28.3    & 88.6    & 97.4     & 95.0     & 37.6     & 58.7    \\
 MAE-AST Frame     & 6        & Random  & 25.9    & 88.0    & 97.8     & 96.6     & 58.6     & 60.2    \\
 MAE-AST Patch     & 12       & Chunked & \textbf{30.6}    & \textbf{90.0}    & 97.9     & 95.8     & \multicolumn{0}{c}{-}   & 59.8    \\
 MAE-AST Frame     & 12       & Random  & 23.0    & 88.9    & 98.0     & \textbf{97.3}     & 63.3     & \textbf{62.1}    \\
    \bottomrule
  \end{tabular}
  
\end{table*}

\begin{table}[th]
  \caption{Decoder Layer Ablation. Based on the default model with solely the decoder depth changed during pretraining}
  \label{tab:decoder_ablation}
  \centering
  \begin{tabular}{l c c c c c c}
    \toprule
    \multicolumn{1}{l}{\textbf{Layers}} & 
    \multicolumn{0}{c}{\textbf{AS}} & 
    \multicolumn{0}{c}{\textbf{ESC}} & 
    \multicolumn{0}{c}{\textbf{KS2}} & 
    \multicolumn{0}{c}{\textbf{KS1}} & 
    \multicolumn{0}{c}{\textbf{SID}} & 
    \multicolumn{0}{c}{\textbf{ER}} \\
    \midrule
1            & \textbf{28.8}    & 87.8      & 97.3   & \textbf{95.5}   & \textbf{42.8}   & 57.8  \\
2            & 28.3    & 88.6    & \textbf{97.4}     & 95.0     & 37.6     & 58.7    \\
3            & 27.8    & \textbf{88.7}    & \textbf{97.4}     & 94.5     & 38.7     & 58.1    \\
4            & 27.6    & 88.1    & 97.2     & 94.4     & 32.7     & \textbf{59.1}         \\
    \bottomrule
  \end{tabular}
  
\end{table}

\textbf{Impact of the Number of Decoder Layers.} In table \ref{tab:decoder_ablation}, we show the downstream results from pretraining the default MAE-AST model with four different decoder depths, all other factors held constant. Like the MAE paper demonstrated for the visual domain, we find that increasing the decoder depth only provides small improvements if any, meaning that the decoder depth can be small relative to the encoder. This result is key in justifying the MAE-AST as an expandable architecture, as decoder layers are significantly more expensive than encoder layers. Interestingly, different tasks show different trends relative to the number of decoder layers, with AS and SID improving with fewer decoder layers and ER improving with more.

\begin{table}[th]
  \caption{Masking Strategies Ablation. Results based on the default model. Tests the pairwise combinations of tokenization type, the masking ratio (\%p), and the masking strategy.}
  \label{tab:masking_strategies}
  \centering
  \begin{tabular}{l c l c c c}
    \toprule
    \multicolumn{1}{l}{\textbf{Input}} & 
    \multicolumn{1}{c}{\textbf{\%p}} & 
    \multicolumn{1}{l}{\textbf{Masking}} & 
    \multicolumn{0}{c}{\textbf{AS}} & 
    \multicolumn{0}{c}{\textbf{KS2}} & 
    \multicolumn{0}{c}{\textbf{ER}} \\
    \midrule
Patch & 50    & Random  & 27.6    & 97.4     & 57.3    \\
Patch & 50    & Chunk   & 27.8    & 97.4     & 58.2    \\
Patch & 75    & Random  & 26.7    & 97.4     & 58.6    \\
Patch & 75    & Chunk   & \textbf{28.3}    & 97.4     & 58.7    \\
Frame & 50    & Random  & 23.6    & 97.8     & 59.1    \\
Frame & 50    & Chunk   & 24.9    & \textbf{97.9}     & \textbf{62.1}    \\
Frame & 75    & Random  & 25.9    & 97.8     & 60.2    \\
Frame & 75    & Chunk   & 23.4    & \textbf{97.9}     & 61.8    \\
    \bottomrule
  \end{tabular}
  
\end{table}


\textbf{Impact of Masking Strategies.} In table \ref{tab:masking_strategies} we show results of eight main strategies of tokenizing and masking. Like the SSAST paper, we see that the most important factor for training is input tokenization, finding patch-based models better for audio tasks, and frame-based models better for speech tasks. We also observe that chunking for a patch-based MAE-AST improves downstream performance across the board, while this did not occur for the vision-based MAE. This could imply that our audio features are more correlated than MAE's input images. Finally, we find that high masking ratios and chunking provide a consistent benefit for patch-based models, while for frame-based models the results are more varied.

\begin{table}[th]
  \caption{Pretraining Speed Ablation. Speed performance is tested for all models using a batch size of 32 and input audio length truncated to 10 seconds on a single Quadro RTX 8000. MAE-AST models use 2 decoder layers, which show high-quality results in table \ref{tab:decoder_ablation} and maximal fine-tune performance in the MAE paper \cite{MaskedAutoencoders2021}. We reimplement SSAST-like training with a decoder-only MAE-AST (w/ [M]) to account for significant library differences and highlight speed differences solely from the switch to an encoder-decoder architecture.}
  
  \label{tab:pretrain_speed}
  \centering
  \begin{tabular}{l c c c c}
    \toprule
    \multicolumn{0}{l}{\textbf{Mask}} & 
    \multicolumn{0}{c}{\textbf{Lay.}} & 
    \multicolumn{0}{c}{\textbf{Sec/Ep}} & 
    \multicolumn{0}{c}{\textbf{Mem. (MiB)}} & 
    \multicolumn{0}{c}{\textbf{Speedup}} \\
    \midrule
75\%     & 6        & 7016        & 6291           & \multicolumn{1}{c}{$1.80\times$} \\
75\%     & 12       & 8299        & 8227           & \multicolumn{1}{c}{\textbf{2.96$\times$}} \\
\midrule
50\%     & 12       & 12222       & 11639          & \multicolumn{1}{c}{$2.01\times$} \\
    \midrule
w/ [M]   & 6        & 12636       & 9871           & -      \\
w/ [M]   & 12       & 24577       & 17719          & -      \\
    \midrule
SSAST~\cite{gong2021ssast} & 12       & 150755      & 41547          & -      \\
    \bottomrule
  \end{tabular}
  
\end{table}

\textbf{Pretraining Speed improvements.} In table \ref{tab:pretrain_speed}, we evaluate the DNN time per epoch and CUDA memory usage for pretraining the MAE-AST with different numbers of encoder layers and masking ratios. We find that the default SSAST implementation\footnote{\href{https://github.com/YuanGongND/ssast}{https://github.com/YuanGongND/ssast}} is significantly slower than our implementation using fairseq\footnote{\href{https://github.com/pytorch/fairseq}{https://github.com/pytorch/fairseq}}, and so we additionally compare against the MAE-AST with 0 encoder layers and $n=6,12$ decoder layers to better represent the speedup from incorporating MAE-style pretraining. We denote this training with mask tokens included at every layer as ``w/ [M]''. MAE-AST achieves a $3\times$ speedup over pretraining with mask tokens included and a $2\times$ memory reduction, showing a significant and practical improvement in training ordinarily-sized models with modern hardware and efficient libraries. With MAE-AST also converging in the same number of iterations as SSAST (figure \ref{fig:convergence_chart}), we see promising results that the MAE-AST is both efficient and expandable.

\begin{table}[th]
  \caption{Pretext Task Ablation. Evaluates a Generative only or Discriminative only loss. We compare the default patch-based SSAST to the default MAE-AST with retrained AudioSet}
  \label{tab:pretext_task}
  \centering
  \begin{tabular}{l c c c c c c}
    \toprule
    \multicolumn{1}{l}{\textbf{Model}} & 
    \multicolumn{0}{c}{\textbf{AS}} & 
    \multicolumn{0}{c}{\textbf{ESC}} & 
    \multicolumn{0}{c}{\textbf{KS2}} & 
    \multicolumn{0}{c}{\textbf{KS1}} & 
    \multicolumn{0}{c}{\textbf{SID}} & 
    \multicolumn{0}{c}{\textbf{ER}} \\
    \midrule
SSAST Gen       & 12.2     & 74.2     & 96.6     & 93.3     & 40.1     & 54.3    \\
SSAST Disc   & \textbf{29.6}     & \textbf{85.6}     & \textbf{98.0}     & \textbf{94.2}     & \textbf{61.4}     & \textbf{57.5}    \\
\midrule
Ours Gen     & 26.1     & 86.3     & \textbf{97.3}     & \textbf{95.1}     & \textbf{40.6}     & \textbf{58.6}    \\
Ours Disc & \textbf{28.5}     & \textbf{89.0}     & \textbf{97.3}     & 94.6     & 38.8     & 58.3    \\
    \bottomrule
  \end{tabular}
  
\end{table}

\textbf{Impact of Pretext Task.} Table \ref{tab:pretext_task} shows the downstream performance of SSAST and default MAE-AST when only using the generative or discriminative losses. We find that the MAE-AST is significantly better than the SSAST when using only a generative loss. Alongside weak results on the low level SID task, we hypothesize this is because the information bottleneck in MAE-AST conditions the model to further extract abstract data, similar to embedding spaces in standard Autoencoders. Our results also suggest that the discriminative loss is more useful for audio tasks while the generative loss is more useful for speech tasks.

\section{Conclusion}

We present the MAE-AST, a new model for self-supervised pretraining in the audio domain. We find our model significantly speeds up pretraining, reduces memory costs, and matches or beats the SSAST on the majority of downstream tasks. Like SSAST, we continue to observe that patch-based tokenization is better for audio tasks and frame-based tokenization is better for speech tasks. Furthermore, we find that MAE-like pretraining is more robust to different loss functions, greatly outperforming the SSAST when only using a generative task, but is significantly worse at low-level tasks like speaker identification. We hypothesize that this implies the MAE architecture is encouraging additional abstraction. We hope that future work can build on these findings toward the goal of constructing a scalable multi-domain strategy for self-supervised pretraining.


\bibliographystyle{IEEEtran}

\bibliography{template}

\begin{thebibliography}{10}
\providecommand{\url}[1]{#1}
\csname url@samestyle\endcsname
\providecommand{\newblock}{\relax}
\providecommand{\bibinfo}[2]{#2}
\providecommand{\BIBentrySTDinterwordspacing}{\spaceskip=0pt\relax}
\providecommand{\BIBentryALTinterwordstretchfactor}{4}
\providecommand{\BIBentryALTinterwordspacing}{\spaceskip=\fontdimen2\font plus
\BIBentryALTinterwordstretchfactor\fontdimen3\font minus
  \fontdimen4\font\relax}
\providecommand{\BIBforeignlanguage}[2]{{%
\expandafter\ifx\csname l@#1\endcsname\relax
\typeout{** WARNING: IEEEtran.bst: No hyphenation pattern has been}%
\typeout{** loaded for the language `#1'. Using the pattern for}%
\typeout{** the default language instead.}%
\else
\language=\csname l@#1\endcsname
\fi
#2}}
\providecommand{\BIBdecl}{\relax}
\BIBdecl

\bibitem{devlin2018bert}
J.~Devlin, M.-W. Chang, K.~Lee, and K.~Toutanova, ``Bert: Pre-training of deep
  bidirectional transformers for language understanding,'' \emph{arXiv preprint
  arXiv:1810.04805}, 2018.

\bibitem{radford2019language}
A.~Radford, J.~Wu, R.~Child, D.~Luan, D.~Amodei, and I.~Sutskever, ``Language
  models are unsupervised multitask learners,'' 2019.

\bibitem{brown2020language}
T.~B. Brown, B.~Mann, N.~Ryder, M.~Subbiah, J.~Kaplan, P.~Dhariwal,
  A.~Neelakantan, P.~Shyam, G.~Sastry, A.~Askell, S.~Agarwal, A.~Herbert-Voss,
  G.~Krueger, T.~Henighan, R.~Child, A.~Ramesh, D.~M. Ziegler, J.~Wu,
  C.~Winter, C.~Hesse, M.~Chen, E.~Sigler, M.~Litwin, S.~Gray, B.~Chess,
  J.~Clark, C.~Berner, S.~McCandlish, A.~Radford, I.~Sutskever, and D.~Amodei,
  ``Language models are few-shot learners,'' 2020.

\bibitem{baevski2020wav2vec}
A.~Baevski, Y.~Zhou, A.~Mohamed, and M.~Auli, ``wav2vec 2.0: A framework for
  self-supervised learning of speech representations,'' \emph{Advances in
  Neural Information Processing Systems}, vol.~33, 2020.

\bibitem{hsu2021hubert}
W.-N. Hsu, Y.-H.~H. Tsai, B.~Bolte, R.~Salakhutdinov, and A.~Mohamed, ``Hubert:
  How much can a bad teacher benefit asr pre-training?'' in \emph{ICASSP
  2021-2021 IEEE International Conference on Acoustics, Speech and Signal
  Processing (ICASSP)}.\hskip 1em plus 0.5em minus 0.4em\relax IEEE, 2021, pp.
  6533--6537.

\bibitem{DBLP:journals/corr/abs-2010-11929}
\BIBentryALTinterwordspacing
A.~Dosovitskiy, L.~Beyer, A.~Kolesnikov, D.~Weissenborn, X.~Zhai,
  T.~Unterthiner, M.~Dehghani, M.~Minderer, G.~Heigold, S.~Gelly, J.~Uszkoreit,
  and N.~Houlsby, ``An image is worth 16x16 words: Transformers for image
  recognition at scale,'' \emph{CoRR}, vol. abs/2010.11929, 2020. [Online].
  Available: \url{https://arxiv.org/abs/2010.11929}
\BIBentrySTDinterwordspacing

\bibitem{liu2021Swin}
Z.~Liu, Y.~Lin, Y.~Cao, H.~Hu, Y.~Wei, Z.~Zhang, S.~Lin, and B.~Guo, ``Swin
  transformer: Hierarchical vision transformer using shifted windows,'' in
  \emph{Proceedings of the IEEE/CVF International Conference on Computer Vision
  (ICCV)}, 2021.

\bibitem{gong2021ssast}
Y.~Gong, C.-I.~J. Lai, Y.-A. Chung, and J.~Glass, ``Ssast: Self-supervised
  audio spectrogram transformer,'' \emph{arXiv preprint arXiv:2110.09784},
  2021.

\bibitem{gong21b_interspeech}
Y.~Gong, Y.-A. Chung, and J.~Glass, ``{AST: Audio Spectrogram Transformer},''
  in \emph{Interspeech}, 2021.

\bibitem{DBLP:journals/corr/VaswaniSPUJGKP17}
\BIBentryALTinterwordspacing
A.~Vaswani, N.~Shazeer, N.~Parmar, J.~Uszkoreit, L.~Jones, A.~N. Gomez,
  L.~Kaiser, and I.~Polosukhin, ``Attention is all you need,'' \emph{CoRR},
  vol. abs/1706.03762, 2017. [Online]. Available:
  \url{http://arxiv.org/abs/1706.03762}
\BIBentrySTDinterwordspacing

\bibitem{Tay2020EfficientTA}
Y.~Tay, M.~Dehghani, D.~Bahri, and D.~Metzler, ``Efficient transformers: A
  survey,'' \emph{ArXiv}, vol. abs/2009.06732, 2020.

\bibitem{MaskedAutoencoders2021}
K.~He, X.~Chen, S.~Xie, Y.~Li, P.~Doll{\'a}r, and R.~Girshick, ``Masked
  autoencoders are scalable vision learners,'' \emph{arXiv:2111.06377}, 2021.

\bibitem{ott2019fairseq}
M.~Ott, S.~Edunov, A.~Baevski, A.~Fan, S.~Gross, N.~Ng, D.~Grangier, and
  M.~Auli, ``fairseq: A fast, extensible toolkit for sequence modeling,'' in
  \emph{Proceedings of NAACL-HLT 2019: Demonstrations}, 2019.

\bibitem{adam}
D.~P. Kingma and J.~Ba, ``Adam: {A} method for stochastic optimization,'' in
  \emph{ICLR}, 2015.

\bibitem{oord2018representation}
A.~v.~d. Oord, Y.~Li, and O.~Vinyals, ``Representation learning with
  contrastive predictive coding,'' \emph{arXiv preprint arXiv:1807.03748},
  2018.

\bibitem{gemmeke2017audio}
J.~F. Gemmeke, D.~P. Ellis, D.~Freedman, A.~Jansen, W.~Lawrence, R.~C. Moore,
  M.~Plakal, and M.~Ritter, ``{Audio Set}: An ontology and human-labeled
  dataset for audio events,'' in \emph{ICASSP}, 2017.

\bibitem{panayotov2015librispeech}
V.~Panayotov, G.~Chen, D.~Povey, and S.~Khudanpur, ``Librispeech: an asr corpus
  based on public domain audio books,'' in \emph{ICASSP}, 2015.

\bibitem{piczak2015esc}
K.~J. Piczak, ``{ESC}: Dataset for environmental sound classification,'' in
  \emph{Multimedia}, 2015.

\bibitem{warden2018speech}
P.~Warden, ``Speech commands: A dataset for limited-vocabulary speech
  recognition,'' \emph{arXiv preprint arXiv:1804.03209}, 2018.

\bibitem{nagrani2020voxceleb}
A.~Nagrani, J.~S. Chung, W.~Xie, and A.~Zisserman, ``Voxceleb: Large-scale
  speaker verification in the wild,'' \emph{Computer Speech \& Language},
  vol.~60, p. 101027, 2020.

\bibitem{busso2008iemocap}
C.~Busso, M.~Bulut, C.-C. Lee, A.~Kazemzadeh, E.~Mower, S.~Kim, J.~N. Chang,
  S.~Lee, and S.~S. Narayanan, ``Iemocap: Interactive emotional dyadic motion
  capture database,'' \emph{Language resources and evaluation}, vol.~42, no.~4,
  pp. 335--359, 2008.

\bibitem{yang2021superb}
S.-w. Yang, P.-H. Chi, Y.-S. Chuang, C.-I.~J. Lai, K.~Lakhotia, Y.~Y. Lin,
  A.~T. Liu, J.~Shi, X.~Chang, G.-T. Lin \emph{et~al.}, ``Superb: Speech
  processing universal performance benchmark,'' \emph{arXiv preprint
  arXiv:2105.01051}, 2021.

\end{thebibliography}

\end{document}